# ANISOTROPY OF CONDUCTIVITY IN BILAYER GRAPHENE WITH RELATIVELY SHIFTED LAYERS

**V.G. LITOVCHENKO, A.I. KURCHAK, M.V. STRIKHA**

V.E. Lashkaryov Institute of Semiconductor Physics, Nat. Acad. of Sci. of Ukraine
*(41, Nauky Ave., Kyiv 03028, Ukraine; e-mail: maksym_strikha@hotmail.com)*

*A transformation of the band structure in bilayer graphene (BLG) with relatively shifted layers has been studied in the framework of the tight-binding model. BLG is demonstrated to remain a zero-gap material in the whole range of experimentally attainable shifts, but the positions of contact points between the conduction and valence bands depend substantially on the shift direction. The shift results in a considerable anisotropy of the band spectrum, which is, in turn, responsible for a substantial (10–20%) anisotropy of the conductivity in BLG. A possibility of using this anisotropy in high-sensitive sensors of a mechanical tension and for the generation of a purely valley current in multivalley anisotropic BLG in the case where both the average spin and the average current of electrons are equal to zero is discussed.*

## 1. Introduction

Bilayer graphene (BLG) has been intensively studied for the last years (see review [1] and references therein). It consists of two graphene layers, which form, owing to the requirement of energy minimization, a configuration of the so-called A-B (Bernal) packing (Fig. 1, *a*), when a half of the atoms in the "upper" layer are located over the atoms belonging to the "lower" one. Actually, BLG is an intermediate structure between single-layer graphene and bulk graphite. BLG, as well as single-layer graphene, is a zero-gap material. However, in a vicinity of point K in the Brillouin zone, its spectrum is not linear, but quadratic (within the energy interval of meVs and tens of meVs, it changes its functional dependence several times and ultimately becomes linear, as the quasiwave vector in the *xy* plane increases) [2].

The interest in studying BLG was strengthened by the phenomenon revealed almost at once and consisting in that, when an electric field is applied along the axis *z*, the difference that emerges between the electrochemical potentials in the "upper" and "lower" layers gives rise to the appearance of a gap between the electron and hole states, as well as to the formation of "Mexican hat" in the energy band spectrum [3]. Since BLG, as well as single-layer graphene, is "doped" – as a rule, by applying some voltage to the gate – this voltage, if one takes a distance of 0.34 nm between the graphene layers in BLG into account, makes those two planes potentially different, with a concentration of $10^{12}$ cm$^{-2}$ corresponding to a gap of an order of 10 meV. This implies that this difference can be neglected in the case of concentrations of an order of $10^{11}$ cm$^{-2}$.

Another way to open the gap consists either in the application of a uniaxial elastic stress, which reduces, similarly to an electric field, the A-B symmetry of BLG and eliminates the degeneration at points K [4, 5], or in the rotation of either of BLG planes with respect to another one [6,7]. In work [8], it was shown within the tight-binding method that the application of a uniaxial stress, within the 6-% limit, to BLG in the $xy$-plane along the "armchair" and "zigzag" directions gives rise to a substantial reconstruction of states near point K; however, no gap between the conduction and valence states arises in this case. In work [9], it was shown by calculations from the first principles that the band-gap opening can be stimulated by applying a mechanical stress to BLG in the direction perpendicularly to the BLG plane. Then the gap remains to be "direct" until the interplane distance exceeds 0.25 nm; whereas the material becomes indirect-gap at higher stresses and, so, smaller distances.

The conductivity and the scattering of charge carriers in BLG have been considered in a significant number of papers (see, e.g., works [10–13]). The consideration was carried out both within the Boltzmann approximation and with the use of more complicated numerical models. However, since the times of scattering by charged impurities in the substrate and by short-range inhomogeneities in BLG itself seem to be of the same order for real structures, and the problem of screening in a 2D structure is difficult, the theoretical description of those processes is far from completeness.

In this work, we analyze a situation, the experimental realization of which can be simpler than those proposed in works [4–7]. Within the tight-binding model, we consider a relative shift of two "unstrained" layers in BLG with respect to each other, which can be described by the angle $\theta$ (see Fig. 1, b). This arrangement can be obtained by putting BLG between two dielectric substrates with opposite bias voltages applied to them. In so doing, the distance between the BLG layers is adopted to be constant and is equal 3.4 °A. This value imposes a restriction on the magnitude of possible shift, which should be smaller than the atomic radius (0.8 °A for carbon), and, hence, on the shift angle $\theta \leq 6°$.

The structure of this work is as follows. In Section 2, the band structure in BLG with the layers relatively shifted with respect to each other is calculated within the tight-binding method. In Section 3, in the framework of the standard calculation scheme of the conductivity in multivalley materials, we demonstrate that such a shift results in the appearance of a conductivity anisotropy. The results obtained and the used approximations are discussed in Conclusions.

## 2. Band Structure in BLG with Shifted Graphene Layers

Within the tight-binding method and analogously to what was done in work [8], the BLG wave function $\psi$ is constructed as a linear combination of wave functions $\chi$ centered at four neighboring atoms with coordinates $\mathbf{r}_{A, B}$ in two planes (see Fig. 2),

$$\psi = C_{A1}\frac{1}{\sqrt{N}}\sum_{A1} e^{i\vec{k}\vec{r}_{A1}}\chi(\vec{r}-\vec{r}_{A1}) + C_{B1}\frac{1}{\sqrt{N}}\sum_{B1} e^{i\vec{k}\vec{r}_{B1}}\chi(\vec{r}-\vec{r}_{B1})$$
$$+ C_{A2}\frac{1}{\sqrt{N}}\sum_{A2} e^{i\vec{k}\vec{r}_{A2}}\chi(\vec{r}-\vec{r}_{A2}) + C_{B2}\frac{1}{\sqrt{N}}\sum_{B2} e^{i\vec{k}\vec{r}_{B2}}\chi(\vec{r}-\vec{r}_{B2})$$
(1)

Plane 2 can be shifted with respect to plane 1 in any direction. The corresponding shift is described as a linear shift of this plane,

$$\delta x = I_c \tan[\theta]\cos[\varphi],\tag{2}$$
$$\delta y = I_c \tan[\theta]\sin[\varphi],\tag{3}$$

where $I_c = 0.34$ nm is the distance between the BLG planes, the angle $\theta$ is shown in Fig. 1,b, and $\varphi$ is the angle between the shift direction and the axis $x$.

In view of Eqs. (2) and (3), the BLG Hamiltonian [2, 8] is modified as follows:

$$H = \begin{pmatrix} \gamma_6 & \gamma_0(h_1 + h_2^*) & \gamma_1 & \gamma_4'h_1^* + \gamma_4''h_2 \\ \gamma_0(h_1^* + h_2) & 0 & \gamma_4'h_1^* + \gamma_4''h_2 & \gamma_3'h_1 + \gamma_3''h_2^* \\ \gamma_1' & \gamma_4'h_1 + \gamma_4''h_2^* & \gamma_6 & \gamma_0(h_1^* + h_2) \\ \gamma_4'h_1 + \gamma_4''h_2^* & \gamma_3'h_1^* + \gamma_3''h_2 & \gamma_0(h_1 + h_2^*) & 0 \end{pmatrix}$$
(4)

where

$$h_1 = e^{ik_x[b+\delta x]},\tag{5}$$

$$h_2 = 2\cos\left(k_y\left[\frac{b\sqrt{3}}{2}+\delta y\right]\right)e^{ik_x\left[\frac{b}{2}+\delta x\right]},\tag{6}$$

and $b = 0.142$ nm is the length of the bond between two atoms in the graphene plane. In addition, the overlap integrals between the neighboring atoms depicted in Fig. 2 were modified with regard for the shift described by Eqs. (2) and (3), so that

$$\gamma_1' = \frac{\gamma_1 I_c^2}{I_c^2 + \delta x^2 + \delta y^2},\tag{7}$$

$$\gamma_3' = \frac{\gamma_3 \left(b^2 + I_c^2\right)}{(b+\delta x)^2 + I_c^2 + \delta y^2}, \tag{8}$$

$$\gamma_3'' = \frac{\gamma_3 \left(b^2 + I_c^2\right)}{\left(\frac{b}{2}+\delta x\right)^2 + \left(\frac{b\sqrt{3}}{2}+\delta y\right)^2 + I_c^2}, \tag{9}$$

$$\gamma_4' = \frac{\gamma_4 \left(b^2 + I_c^2\right)}{(b+\delta x)^2 + I_c^2 + \delta y^2}, \tag{10}$$

$$\gamma_4'' = \frac{\gamma_4 \left(b^2 + I_c^2\right)}{\left(\frac{b}{2}+\delta x\right)^2 + \left(\frac{b\sqrt{3}}{2}+\delta y\right)^2 + I_c^2}. \tag{11}$$

In what follows, we use the standard numerical values of overlap integrals in BLG without deformation [8]; namely, $\gamma_0 = 2.598$ eV describes the binding energy between neighboring atoms in the same graphene plane, $\gamma_1 = 0.364$ eV describes the binding energy between two atoms A from different planes, $\gamma_3 = 0.319$ eV describes the binding energy between two atoms B from different planes, and $\gamma_4 = 0.177$ eV describes the binding energy between atom A from one plane and atom B from another plane (see Fig. 2). Making allowance for different chemical environments of atoms in positions A and B results in the appearanceof the small integral $\gamma_6 = -0.026$ eV.

The numerical solution of the eigenvalue problem (4) gives the known structure of the BLG band spectrum (see Fig. 3, where, for illustrative purposes, only one of two doublets of values is shown; namely, this is the doublet that forms the conduction and valence bands). In this case, for all shift values $\theta < 6°$, i.e. when our model is still valid, BLG remains a material with the zero energy gap. However, the locations of contact points between the conduction and valence bands considerably depends on the direction of a relative plane shift (Fig. 4). This is an evident consequence of the symmetry reduction under the action of the relative plane shift in the system including a lattice with a basis. Another consequence of the shift is the appearance of a substantial band anisotropy. It should be noted that a similar situation takes place for single-layer graphene; namely, the material remains zero-gap up to large values (10–20%) of an induced strain, but the anisotropy emerges in it as well (see, e.g., work [14] and references therein).

In Fig. 5, the band spectra of undeformed BLG and BLG with the planes shifted with respect to each other along the axis $x$ are exhibited. They were calculated in a vicinity of the upper of six extrema depicted in Fig. 4. One can see that the anisotropy (in the English language journals, the term "warping" is used) in undeformed BLG for the selected directions is insignificant up to energies of about 0.5 eV. However, provided that the planes are shifted, it becomes substantial. In particular, the effective mass remains almost invariant along the semimajor axis and considerably decreases along the semiminor one.

Note that, generally speaking, the effective mass approximation should be applied to BLG with great care everywhere, but in a close vicinity (of a width of an order of 1 meV) of point K because of the reasons explained in work [2]. However, since the real spectra obtained as numerical solutions with Hamiltonian (4) can be approximated by effective mass ellipses in the plane ($k_x$, $k_y$) with an accuracy of about 5%, it is this approximation that will be used below.

Similar results for the shift along the axis $y$ and along the direction at 45° with respect to the axis $x$ are shown in Figs. 6 and 7, respectively. One can see from Fig. 6 that the results obtained for the shift along the axis $y$ are qualitatively similar, but the semimajor and semiminor axes of the ellipse interchange.

On the other hand, for the shift along the direction at 45° with respect to the axis $x$ (Fig. 7), the masses also become anisotropic; however, their values are much smaller than in the undeformed case. At the same time, we should note that the shift reduces the symmetry of the problem and splits six physically equivalent extrema (Fig. 4, *a*) into two groups consisting of two and four equivalent extrema (Figs. 4, *b* to *d*). Hence, those groups are characterized by their own effective masses, which will be denoted by subscripts 1 and 2. The mass values were calculated in the simple parabolic approximation of the dispersion law and its exact anisotropic form obtained numerically for an energy of 0.09 eV, which corresponds to the actual concentration values of an order of $10^{11}$ cm$^{-2}$ in BLG. The specific values obtained are listed in Table.

## 3. Anisotropy of Conductivity in BLG with Shifted Graphene Layers

Let us consider the electric conductivity in BLG with shifted graphene layers in the framework of the standard scheme applied to multivalley materials (see, e.g., work [15]). If the semiconductor has $M$ valleys in the conduction band, the total current density is equal to the sum of current densities over the valleys,

$$j = \sum_{v=1}^{M} j^{(v)} \quad (12)$$

In order to find the current density for the $v$-th valley, the electric field vector **E** should be written down in the coordinate system connected with the principal axes of the effective mass tensor for this valley. Since the problem is two-dimensional (2D), the vector **E** looks like

$$E = \left( E_1^{(v)}, E_2^{(v)} \right) \quad (13)$$

Then, the sought current reads

$$j^{(v)} = \left(j_1^{(v)}, j_2^{(v)}\right) = \left(\sigma_1^{(v)} E_1^{(v)}, \sigma_2^{(v)} E_2^{(v)}\right) \quad (14)$$

In the consideration below, we use the standard "Drude" expression for the specific conductivity, taking into account that, in the 2D case, the concentration of charge carriers has a dimension of m$^{-2}$, and the specific conductivity a dimension of $\Omega$ (see, e.g., work [16]),

$$\sigma_i^{(v)} = \frac{e^2 n^{(v)}}{m_i} \langle \tau_\rho \rangle \quad (15)$$

where $n^{(v)}$ is the concentration of charge carriers in the $v$–th valley, and $<\tau_p>$ is the average relaxation time, the possible anisotropy of which will be discussed below.

The current $j$ can be expressed in the form, in which all conductivity tensors $\sigma^{(v)}$ are written down in the same coordinate frame. Then

$$j = \sum_{v=1}^{M} \sigma^{(v)} E = \sigma E \quad (16)$$

where the tensor $\sigma$ is the sum of tensors $\sigma^{(v)}$.

In the case concerned, there are six energy minima. They are associated with two equivalent valleys in undeformed BLG, because only one third of those minima belong to the first Brillouin zone. If the graphene layers are shifted with respect to each other, the equivalence among the valleys becomes broken.

Therefore, we sum up in Eq. (12) over all six minima, by bearing in mind that the final result should be divided by 3 (Fig. 8). Let the isoenergy surfaces be elliptic near the conduction band bottom (the limits of applicability of this approximation were discussed above). First, let us consider each ellipse separately in its own coordinate system and write down the corresponding conductivity tensors, by preliminarily reducing them to their principal axes. With regard for the numeration of minima (Fig. 8) and their splitting into two groups including two and four equivalent minima (Fig. 4) under a shift of BLG layers with respect to each other, we obtain

$$\sigma^{(1)} = \sigma^{(4)} = \begin{pmatrix} \sigma_{11}^{(1)} & 0 \\ 0 & \sigma_{22}^{(1)} \end{pmatrix}, \quad \sigma^{(2)} = \sigma^{(3)} = \sigma^{(5)} = \sigma^{(6)} = \begin{pmatrix} \sigma_{11}^{(2)} & 0 \\ 0 & \sigma_{22}^{(2)} \end{pmatrix} \quad (17)$$

Since the ellipses in each group – (1, 4) and (2, 3, 5, 6) – are equivalent, the corresponding conductivity tensors must also be identical.

The conductivity tensors $\sigma^{(v)}$ have to be written down in the same coordinate system (X, Y, Z). For this purpose, we select the coordinate system connected with the tensors $\sigma^{(1)}$ and $\sigma^{(4)}$. This means that those two tensors remain invariant, whereas the others have to be rewritten in accordance with the rules of tensor component transformation when changing from one coordinate system to another,

$$A'_{ik} = \alpha_{i'l}\alpha_{k'm} A_{lm} \quad , \qquad (18)$$

where $\alpha_{ij}$ are the cosines of angles between the corresponding coordinate axes. The corresponding components of the tensors $\sigma^{(2,3,5,6)}$ in the coordinate system $(X, Y, Z)$ look like

$$\begin{aligned}\sigma'_{11} &= \alpha^2_{1'1}\sigma_{11} + \alpha^2_{1'2}\sigma_{22} \\ \sigma'_{12} &= \alpha_{1'1}\alpha_{2'1}\sigma_{11} + \alpha_{1'2}\alpha_{2'2}\sigma_{22} \\ \sigma'_{21} &= \alpha_{2'1}\alpha_{1'1}\sigma_{11} + \alpha_{2'2}\alpha_{1'2}\sigma_{22} \\ \sigma'_{22} &= \alpha^2_{2'1}\sigma_{11} + \alpha^2_{2'2}\sigma_{22}\end{aligned} \qquad (19)$$

Then, the tensor of total electroconductivity can be written down as follows:

$$\sigma = 2\sigma^{(1,4)} + 4\sigma'^{(2,3,5,6)} \qquad (20)$$

where the components of tensors $\sigma'^{(2,3,5,6)}$ are expressed in terms of the components of the tensors $\sigma^{(2,3,5,6)}$ in view of Eq. (19).

The formula that allows us to determine $\sigma_{ii}$ for each ellipse has the standard form (15). It includes the concentration expressed in terms of the 2D density of states $D(v)(E)$ of the degenerate electron gas as [16]

$$n = \sum_{1,4} \int_0^{E_f} D_{1,4}(E)dE + \sum_{2,3,5,6} \int_0^{E_f} D_{2,3,5,6}(E)dE \quad , \quad (21)$$

where $E_f$ is the Fermi energy. The expression for the 2D density of states obtained with regard for the effective mass anisotropy at energies below the second quantized level [16] and assuming the anisotropic square-law band spectrum looks like

$$D_{(v)} = \frac{2(m^{(v)}{}_l\, m^{(v)}{}_t)^{1/2}}{\pi \hbar^2} \qquad (22)$$

In view of Eqs. (15), (17)–(19), (21), and (22), as well as the effective mass values quoted in Table 1, tensor (20) in the common coordinate system $(X, Y, Z)$ can be written down as follows.

1. If there is no shift,

$$\sigma = \frac{e^2 \langle \tau_p \rangle E_f}{\pi \hbar^2} \begin{pmatrix} 2 & 0 \\ 0 & 2 \end{pmatrix}, \tag{23}$$

and the standard isotropic conductivity in BLG takes place.

2. For a shift along the axis $x$,

$$\sigma = \frac{e^2 \langle \tau_p \rangle E_f}{\pi \hbar^2} \begin{pmatrix} 2.14 & 0 \\ 0 & 1.88 \end{pmatrix}, \tag{24}$$

i.e. there emerges an appreciable anisotropy associated with a shift-induced anisotropy of the band spectrum. The conductivity along the axis $x$ turns out to be higher.

3. For a shift along the axis $y$,

$$\sigma = \frac{e^2 \langle \tau_p \rangle E_f}{\pi \hbar^2} \begin{pmatrix} 1.94 & 0 \\ 0 & 2.07 \end{pmatrix}, \tag{25}$$

i.e. the anisotropy also emerges. But, in this case, the conductivity is higher along the axis $y$.

4. For a shift along a direction at 45° with respect to the axis $x$,

$$\sigma = \frac{e^2 \langle \tau_p \rangle E_f}{3\pi \hbar^2} \begin{pmatrix} 2.05 & 0 \\ 0 & 1.96 \end{pmatrix} \tag{26}$$

Hence, the highest anisotropy of the conductivity (about 10%) arises, when the layers are relatively other directions, the resulting anisotropy is approximately half as high.

Up to this point, we assumed that the relaxation time does not depend on the direction. This assumption needs a separate discussion. In the case where the charge carriers are scattered by charged impurities in a substrate, and the concentration of those impurities is rather low, i.e. when the effective mass approximation is valid, the time $\tau p$ is determined by modified formula (A16) from work [13]:

$$\frac{1}{\tau_p} = \frac{(m_l m_t)^{1/2}}{\hbar} v_f^2 I. \tag{27}$$

Here, $v_f = 10^8$ cm/s, and $I$ is a dimensionless integral depending on the specific scattering potential and its screening, as well as the correlation in the arrangement of such potentials and the relative concentration of scattering impurities in the substrate. According to Eqs. (23)–(26), the additional account for the scattering time anisotropy using formula (27) makes the conductivity anisotropy approximately twice as high.

However, the applicability scope of formula (27) still remains debatable, as well as the general issues concerning the mechanisms of charge carrier scattering in BLG [11–13].

## 4. Conclusions

The transformation of the band spectrum in bilayer graphene (BLG) with relatively shifted graphene layers has been studied within the tight-binding method. BLG is shown to remain a material with the zero energy gap in the whole interval of experimentally attainable layer shifts. However, the locations of contact points between the conduction and valence bands substantially depend on the direction of a shift of the planes with respect to each other. This phenomenon is a consequence of the symmetry reduction in the system including a lattice with a basis under the shift action.

The shift results in the appearance of a considerable band anisotropy, which brings about, in turn, a substantial (about 10–20%) anisotropy of the BLG conductivity. This phenomenon can be applied to high-sensitive sensors of mechanical tension. Another way of its application consists in the generation of a purely valley current in anisotropic multivalley BLG, provided that both the average electron spin and the average electron current are equal to zero (the socalled "valleytronics", see works [17, 18]). A possibility of such generation in semiconductor quantum wells at direct subband and intraband optical transitions was discussed for the first time in work [19].

The mechanism of band "separation" in undeformed graphene with the help of polarized light (with the use of "natural" anisotropy – "warping" – of the band spectrum at high energies corresponding to a transition in the visible range) was proposed in work [20].

One of the possible schemes for the experimental generation of a "valley" current in graphene was described in work [21]. The generation of valley currents in the deformed single-layer graphene was theoretically considered in work [22]. Those effects could be expectedly observed more easily in deformed BLG, because the anisotropy of the band spectrum emerges in this case already at low kinetic energies of charge carriers.

We should emphasize that our results were obtained in the framework of a number of approximations, with their limits of applicability being indicated above. First of all, we neglected the gap opening if the electric voltage was applied along the axis $z$, which put a restriction on the magnitude of graphene "doping" by the gate (to $10^{11}$ cm$^{-2}$). Second, the approximation of elliptic-like bands forced us to consider the charge carriers at kinetic energies higher than 10 meV, at which the bands do not have three lateral minima any more [2]. At last, the approximation of the invariable interplane distance equal to 0.34 nm imposed an evident restriction on the shift magnitude (the angle $\theta$ should not exceed 6◦). At larger deformations (finally, they would result in the $sp^2-sp^3$ hybridization and the appearance of a gap between the conduction and valence bands [23, 24]), the consideration would go beyond the limits of the simple tight-binding approximation [8] used in this work

*The work was sponsored by the State Fund for Fundamental Researches of Ukraine (grant 53.2/006). The authors are grateful to T.V. Linnik for the valuable useful discussion.*

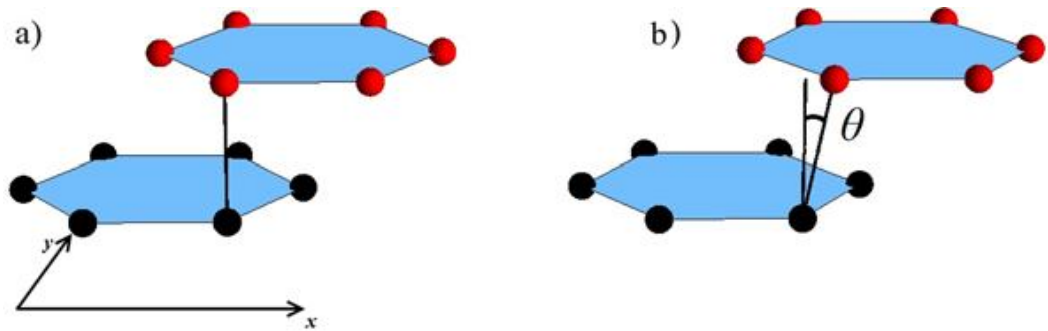

*Fig. 1*. Bernal packing in BLG (*a*) with no relative shift of the graphene layers and (*b*) when the layers are shifted with respect to each other along the axis *x*.

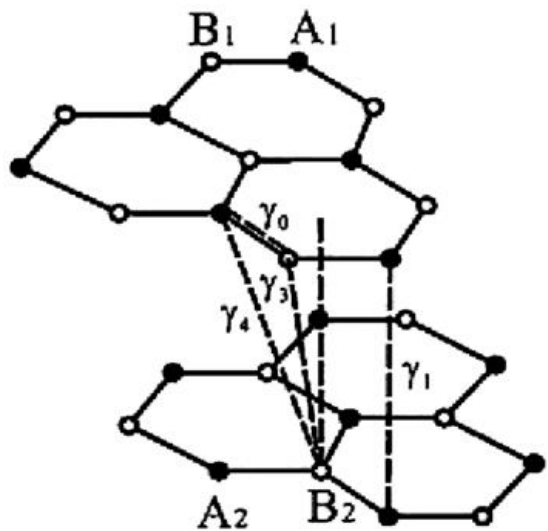

*Fig. 2*. Arrangement of atoms A1,2 and B1,2 in two BLG planes in the "Bernal" packing. Positions A are occupied by atoms in both planes, whereas positions B only in either of them. The axis *x* corresponds to the "armchair" direction, and the axis *y* to the "zigzag" one. Marked are the pairs of closely located atoms that form the corresponding overlap integrals $\gamma$.

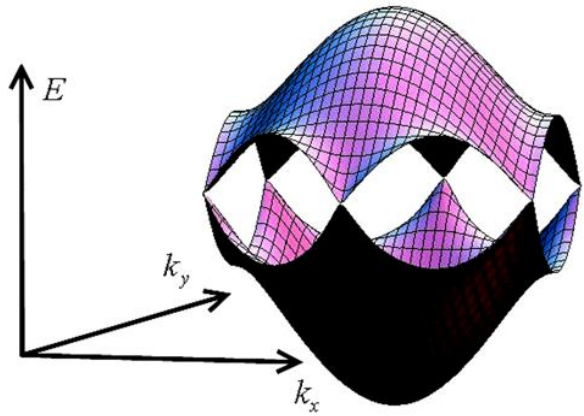

*Fig. 3.* General view of the structure of the "lower" doublet of two BLG doublets

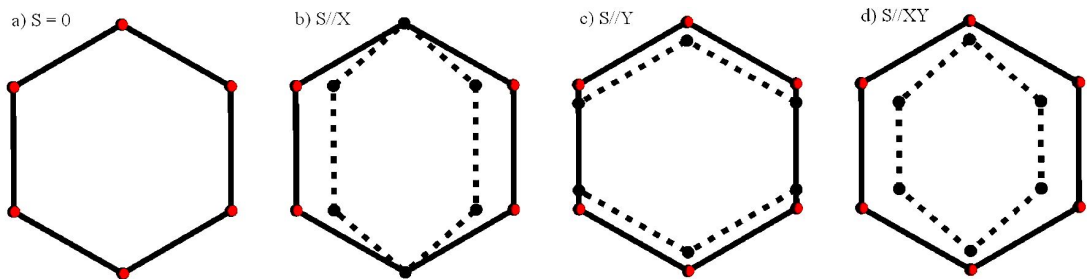

*Fig. 4.* Modifications of the positions of contact points between the conduction and valence bands for various shift directions *S*: (*a*) no shift, (*b*) shift along the axis *x*, (*c*) shift along the axis *y*, and (*d*) shift along a direction at 45° with respect to the axis *x* or *y*

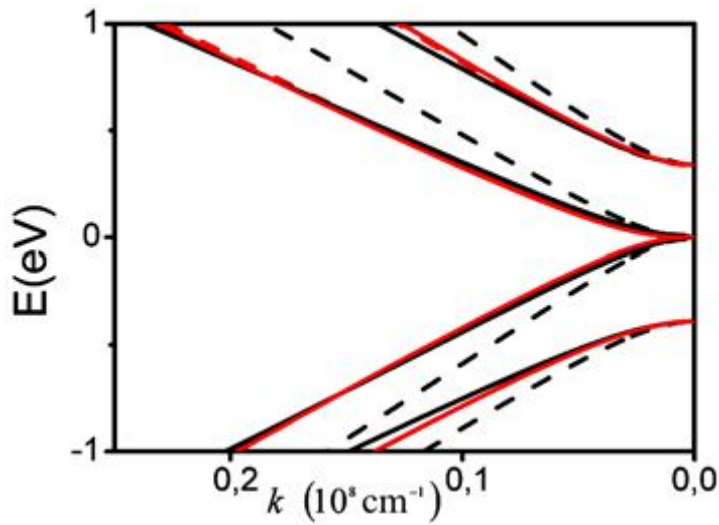

*Fig. 5.* Band spectra of undeformed BLG (solid curves) and BLG with the layers shifted along the axis *x* (dashed curves, calculations for $\theta = 5°$) in a vicinity

of the upper extremum in Fig. 4. Dark (black) curves and points correspond to the direction along $k_x$, light (red) ones to the direction along $k_y$

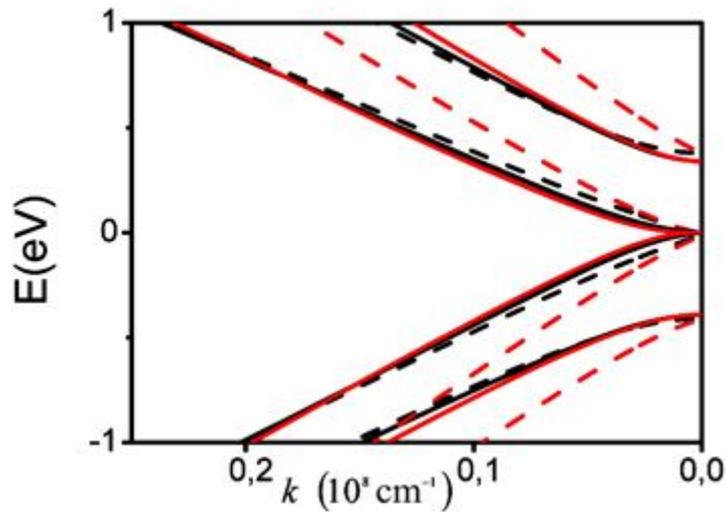

*Fig. 6.* The same as in Fig. 5, but for the shift along the axis *y*

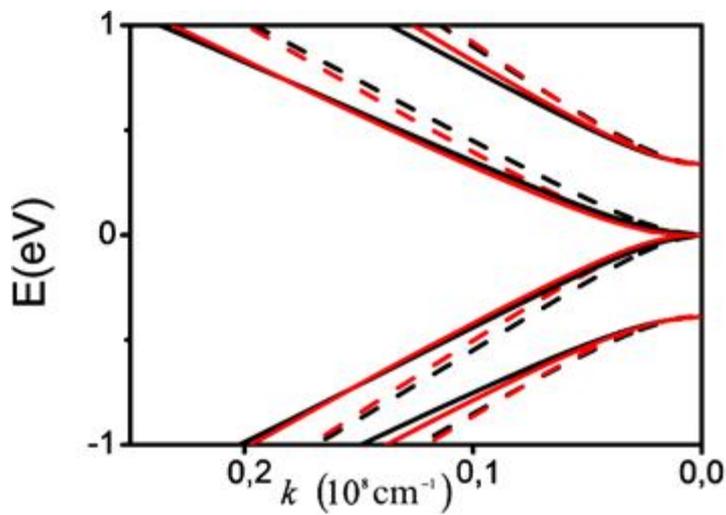

*Fig. 7.* The same as in Fig. 5 for a shift along the direction at 45° with respect to the axis *x*

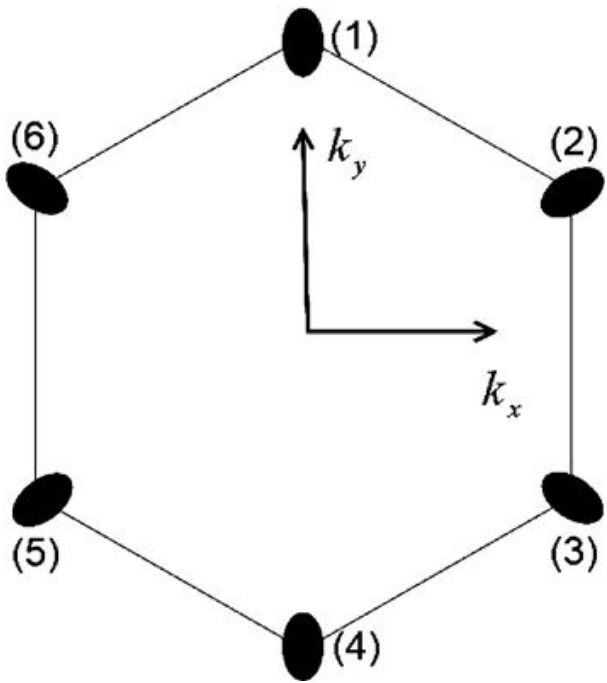

*Fig. 8.* Numeration of minima in deformed BLG

Effective masses (in terms of free electron mass units) in BLG obtained in the approximation of elliptic isoenergy surfaces: with no shift, with the shift along the axis *x*, with the shift along the axis *y*, and with the shift along a direction at 45∘ with respect to the axis *x*. In the last three cases, subscript 1 corresponds to two and subscript 2 to four equivalent extrema exhibited in Fig. 4 subscript 1 corresponds to two and subscript 2 to four equivalent extrema exhibited in Fig. 4

| S = 0 | | S//X | | | | S//Y | | | | S//X Y | | | |
|---|---|---|---|---|---|---|---|---|---|---|---|---|---|
| $m_t$ | $m_l$ | $m_t^1$ | $m_l^1$ | $m_t^2$ | $m_l^2$ | $m_t^1$ | $m_l^1$ | $m_t^2$ | $m_l^2$ | $m_t^1$ | $m_l^1$ | $m_t^2$ | $m_l^2$ |
| 0.064 | 0.069 | 0.042 | 0.062 | 0.051 | 0.051 | 0.051 | 0.053 | 0.048 | 0.058 | 0.041 | 0.053 | 0.044 | 0.050 |

# АНІЗОТРОПІЯ ПРОВІДНОСТІ ДВОШАРОВОГО ГРАФЕНУ ПРИ ВІДНОСНОМУ ЗМІЩЕННІ ЙОГО ШАРІВ

*В.Г. Литовченко, А.І. Курчак, М.В. Стріха*

Р е з ю м е. У рамках методу сильного зв'язку досліджено трансформацію зонного спектра двошарового графену (ДГ) зі зсунутими один щодо одного графеновими шарами. Показано, що в усьому діапазоні експериментально актуальних значень зсуву ДГ залишається матеріалом з нульовою забороненою зоною, однак розташування точок дотикання зони провідності і валентної зони при цьому суттєво залежить від напрямку зсуву площин одна щодо одної. Наслідком зсуву є поява суттєвої анізотропії зон, яка, в свою чергу, приводить до суттєвої (порядку 10–20%) анізотропії провідності ДГ. Обговорено можливе застосування такої анізотропії в чутливих сенсорах механічних напружень та для генерації в анізотропному багатодолинному ДГ суто долинного струму за умови рівності нулеві середнього електронного спіну й електронного струму.